# The Gasdynamics First Problem Solution -
# The Gas Stream Parameters, Structures, Metering Characteristics for Pipe, Nozzle

S.L. Arsenjev, I.B. Lozovitski[1], Y.P. Sirik

*Physical-Technical Group*
*Dobroljubova street 2, 29, Pavlograd, Dnepropetrovsk region, 51400 Ukraine*

The development of the classical physics fundamentals with usage of experience of the experimental and computational-analytical investigations has allowed to create the valid conception of a fluid motion. The multidisciplinary approach based on a new theory has allowed to construct physical adequate mathematical model of the gas flow taking into account essential nonlinearity of change of the gas stream main parameters and the metering characteristics of such flow elements as pipe, pipeline, convergent nozzle and, what is more, with allowance for the evolutional character of the gas flow structure in front of setting, in the flow element itself and in the jet flowing out of it. The example of the valid solution of the direct and inverse problems about a gas flow through the flow element is given here for the first time. The results of solution in amount, kind and precision of the obtained information are an example of the satisfactory solution of the gas dynamics problem.
**PACS:** 05.65.+в; 07.05.Tp; 47.10.+g; 47.27.Wg; 47.40.-x; 47.32.Ff; 47.40.Dc; 47.40.Hg; 47.40.Ki; 47.60.+i

## Introduction

In June 2000 the "VeriGas" computer program was created. For the first time the program is physically adequately reproducing the gas stream dynamics in the flow systems and elements: pipe, pipelines, mouthpieces, nozzles etc. This is the first and key achievement on the way of practical application of the such general physical conception theses as allowing to construct the physically adequate mathematical model of the states and processes in the gas and liquid mediums and deformable body from the unified positions constituting the basis of a new stage of the physics development in contrast to basis of the analogous purpose well-known software products.

Side by side with traditional physics containing the items of information on the Nature separate laws and specific theories and here called the Physics-I, the fundamentals of the physics allowing to describe the phenomena dynamics in the Nature and engineering by means of conjugation of the separate laws into such integral information structure which is rightly and precisely reproducing the state and motion parameters of the contact-interacting real mediums and bodies are created. This is the Physics-II. So:
-    The Physics-I is primary science detecting and accumulating the Nature separate laws and conceiving the separate specific theories;
-    The Physics-II is the second science conceiving the modern working conception of the Nature structure as an integral dynamic system possessing the generality to the diversity of separate phenomena, laws.
Such approach quite corresponds to the Aristotle and Newton guidelines: first gather the facts, and then search for causes of the general. The detection and accumulation of the Nature separate laws is going on and will be going on practically endlessly. However a presence of this process and its results in itself ensures only the necessary but not sufficient condition for creation of the physically adequate model of phenomenon, process. The traditional attempts of description of the real gas, liquid and deformable body state and motion by means of unification of the separate physical laws with the help of mathematics are so unsuccessful during the last 250 years just because of non-compliance with the second condition.

---
[1] Phone: (38 05632) 38892, 40596
E-mail: loz@inbox.ru

As to possibility of creation of the dynamics mathematical model of real phenomenon or process, it is possible only after creation of its general physical conception. Just the conception allows to determine the necessary volume of the separate physical laws in their mathematical expression and shows the way of their conjugation into such an integral information structure which has generality to multitude real variations of phenomena and processes and at the same time ensures the true and exact estimations of the state and motion parameters of material mediums and bodies in every specific case. Such information structure is the only one and it is applied with the help of the **Physical Ensemble** method. The harmonious composing of Physics-I and Physics-II with the help of the Physical Ensemble method is the only way to be out of the protracted crisis state in the scientific-engineering field. Just the **United Physics** is the full value knowledge and at the same time the technological instrument allowing to avoid traditional mathematical scholasticism on one hand and to exclude the industrial experiment or to decrease essentially its scales on the other hand already today. The development of the United Physics and software on its basis will determine the way and scale of modern progress.

The unsatisfactory state of modern conceptions about fluid motion can be demonstrated on an example of a problem about the gas stream motion in the pipe. Pipe is the simplest flow element. Gas has the simplest structure. The simplest formulation on gas flow through the pipe is motion with friction without heat exchange.

The results of the solution of this gasdynamics problem are submitted in this work for the first time. The distinctive feature of the new approach is getting the information about structures of the gas stream motion according to the effective pressure drops, state and motion parameters of the gas stream and metering characteristic of the flow element as well.

The results of solution are an evidence of the complicated evolutional change of the motion structures of the gas stream under the pressure change, essential nonlinearity of the state and motion parameters change and essential nonlinearity of the metering characteristic of simple flow elements. Some facts of this physical concept have been given in papers of authors [1−7].

### General formulation

The simplest formulation of the problem about gas flow through the pipe presupposes a presence of two half-spaces, filled with gas of the identical composition and linked with pipe flow element. Air under normal technical temperature 288 K is taken as gas.

The right half-space is a zone of normal technical absolute pressure $p_h = 10^4$ kgf/m$^2$. The left half-space is a zone of high absolute pressure $p_0 > p_h$. So, the gas flows from left to right. Geometrical parameters of the pipe are: caliber (interior diameter) $D$ (given below), medial relative roughness of the pipe wall $\overline{\Delta}/D = 0.0273$, relative length of the pipe is measured in its calibers $\overline{L} = L/D$ (given below).

According to the simplest formulation, we suppose the gas flow in the flow element to be one-dimensional, stationary and without heat exchange. The friction coefficient (factor) is determined by means of the Colebrook's diagram.

Solving of this problem means determination of state and motion parameters of the gas flow and distinctive geometrical and dynamical parameters in three zones: in front of the pipe - the inflow zone, in the pipe itself and behind the pipe – the outflow zone. This three zones form the General Dynamic Structure of the flow, which ensures physical correctness of the gas flow analysis. The results of computational experiment adduced in this work do not contain the analysis of processes, connected with the normal and oblique shock waves as these processes are well enough described in the educational and methodical literature.

**Computational experiment results**

**1. The evolution of the gas flow structures in the General Dynamic Structure of the flow**
The introduction of concept of the General Dynamic Structure of the flow envisions the analysis of the gas flow from its origin and transformation into a stream in front of the pipe, then the analysis of motion of the gas stream in the flow element and at last, analysis of the gas flow in jet after it's going out of the flow element into atmosphere.
The pipe with interior diameter $D = 134$ mm and relative length $\overline{L} = L/D = 9$ is used in these computations. The rest of requirements correspond to the «General formulation» chapter. The pipe with $\overline{L} = 9$ is taken because of its properties. The consecutive computational experiment has allowed to determine 15 theoretically possible structures of the gas flow; 13 of them (from the 2nd to the 14th) are adduced on fig. 1 under consecutive increase of pressure.

Nomenclature of the gas stream structures on fig. 1:
| | |
|---|---|
| $a$ | high pressure gas half-space – motionless gas; |
| $b$ | inlet zone – acceleration of gas, origin of stream; |
| $c$ | low pressure gas half-space – motionless gas; |
| $d$ | mass flow dispersion zone – multi-jet motion of gas, braking by the gas jets; |
| $e$ | mass flow cone – outflow zone with quasi-permanent parameters; |
| $f_{os}$ | stream with the oblique shock waves in the pipe; |
| $f_{sb}$ | subsonic stream in the pipe; |
| $f_{sbk}$ | subsonic laminar kernel of stream in the pipe; |
| $f_{spk}$ | supersonic laminar kernel of stream in the pipe; |
| $g_{en}$ | vortex sublayer of stream at entrance to the pipe – zone of separation of the stream kernel from the entrance walls of pipe; |
| $g_f$ | vortex sublayer of stream in the pipe – zone of separation of the stream kernel from the pipe wall; |
| $j_{os}$ | jet with oblique shock waves; |
| $j_{sb}$ | expanded subsonic part of jet; |
| $j_{sl}$ | transonic superlayer of the jet supersonic part; |
| $j_{sp}$ | expanded supersonic part of jet; |
| M | Mach number; |
| NSW | normal shock wave. |

Within the framework of the adopted nomenclature, the total $(a+b+c+d+e+f)$ is a legend of the General Dynamic Structure of the flow for such flow element as pipe, nozzle. Other symbols mean the transition structures of the gas flow, which appear and are replaced by the others under pressure change.
One of features of the pipe with relative length equal to 9 calibers is that it is much shorter than the entry length. Therefore transition of the gas stream to the turbulent regime does not take place in it under any pressure drop. The boundary layer thickness accounts rather small share of the section radius of the pipe. The description of structures of the gas stream is built on the basis of description of General Dynamic Structure with the detailed description of the each zone's features.

**The first structure** ($\Delta p \cong 6 \times 10^{-3}$ kgf/m$^2$):
– inflow zone has the shape of a half-sphere in this case according to the **Elementary Wave Theory of the Fluid Motion**; this zone consists from of close and far fields of radiation of the waves by an inflow face of the pipe; the origin of the gas flow and its intensification up to the quantity of the corresponding velocity of the inlet into the pipe, happens as a result of interaction of the overpressure potential $\Delta p$ and the wavefronts of the half-spherical shape closed to the wall of the left half-space and running in the opposite direction to the gas flow;

− gas stream motion in the pipe is going on without acceleration; such a hydraulic pattern of the gas flow is conditioned by small intensity of friction of the stream with the pipe wall and accordingly small heat generation; in result, the gas flow in the pipe is not accompanied by expansion and goes with constant average velocity on all the pipe length; at the same time, the interaction of the gas stream and the pipe wall causes appearance of the wall boundary layer, in which flow velocity is varied from zero up to rated quantity in a thin wall layer;

− gas jet effusing out of the pipe, interacts contactly with the motionless ambient gas medium (atmosphere); the feature of the jet at the outlet of the pipe is that velocity of the surface layers of jet is incremented from zero (as it was in the pipe) to a quantity equal with paraxial velocity under action of viscosity forces in jet; the acceleration distance of the surface layers of jet is commensurable to its diameter and the jet maintains delineations of the pipe on this part of length; the increase of velocity of surface layers causes smoothing of velocity profile in the jet on one hand and intensifies interaction of the jet with the ambient atmosphere on the other hand; intensification of this interaction causes buckling of the contacting layers of the jet and atmosphere and forms ring vortex structures; these structures diffuse the mass flow rate of jet in the ambient air as the result of losing its own ring form stability and forming of isotropic turbulence cloud; visualization of such a jet is given on fig. 117 in the album [8].

**The second structure** ($\Delta p \cong (0.3...1.85) \times 10^4$ kgf/m$^2$):

− inflow zone in the pipe looks like inflow zone of the first structure;

− gas stream motion in the pipe is going on with acceleration up to velocity not exceeding the local sound velocity; acceleration of the stream along the pipe is conditioned by the heat release as a result of friction between the stream and the pipe wall - this is so-called frictional self-acceleration of the gas stream;

− gas jet effusing out of the pipe gets into the ambient air forming so-called mass flow cone; diameter of this cone's base is the same as diameter of the pipe, medial height is ~ 4.5 diameters of its base; the average pressure in the volume of the mass flow cone is the same as ambient air's pressure and the pressure in zone "e" (look at fig. 1) has oscillatory character according to the Elementary Wave Theory of the Fluid Motion and amplitude value of this pressure periodically exceeds pressure of the ambient air; as a result, the mass flow cone periodically pulsewise throws out quantums of the gas flow rate in the kind of mini-jets through its lateral surface into the ambient air; the quantums of the gas flow rate get into ambient air and owing to the presence of radial velocity they carry away the ambient air in the jet motion direction, thanks to the presence of the longitudinal transportation velocity; this is the mass flow dispersion zone "$d$"; the interaction of quantums of the flow rate with the ambient air is finished with formation of the isotropic turbulence cloud.

**The third structure** ($\Delta p \cong 1.88 \times 10^4$ kgf/m$^2$):

− inflow zone in the pipe looks like inflow zone of the first structure;

− gas stream motion is going on with acceleration up to the local sound velocity in the pipe end section;

− gas jet effusing out of the pipe is formed by the oblique shock waves with the high frequency and is terminated by the mass flow cone and mass flow dispersion zone, like the second structure of the jet; the hitting range of this structure jet is much more, than in the first and second structures.

**The fourth structure** ($\Delta p \cong 1.9 \times 10^4$ kgf/m$^2$):

− inflow zone in the pipe looks like inflow zone of the first structure;

− gas stream motion in the pipe is accompanied by appearance of the stream kernel separated from the pipe wall on all its length excepting inlet section; relative velocity M=1 is in front of the pipe outlet section and promptly shifts to inlet of the pipe under small increase of pressure; the nature of the gas flow in the kernel is rather close to isentropic; the nature of the wall sublayer is hydrodynamic down to vortical; thickness of this hydrodynamic sublayer is small according to the computational results and increments from zero in the inlet section of the pipe up to maximum at M=1, and then diminishes the outlet section but not low to zero. Spontaneous forming of a stream kernel with its acceleration from subsonic to sonic and supersonic velocity is so-called molecular self-acceleration of a gas stream. This phenomenon is a nature change of the gas molecules interaction of approaching of a gas stream velocity to its local sonic velocity. Chaotic interaction of gas molecules yields spontaneously to their interaction mainly in direction of the gas stream motion. Molecular self-acceleration is a breach of Clausius-Maxwell's principle because of the velocity resonance of gas stream and of it molecules;

−   gas jet effusing out of the pipe looks like the 3rd structure; the only difference is that a step of the oblique shock waves is a bit more and the jet kernel is surrounded by the transonic superlayer.

**The fifth structure** ($\Delta p \cong 2.4 \times 10^4$ kgf/m$^2$):
−   inflow zone in the pipe looks like inflow zone of the first structure;
−   gas stream motion in the pipe is accompanied by bias of section M=1 to the inlet section of the pipe; the hydrodynamic sublayer is shortening, its thickness incrementing a little to the M=1 section; the kernel of the stream within limits of the sublayer length accelerates up to the supersonic velocity; therefore the stream motion is penetrated by the system of oblique shock waves on the rest part of the pipe;
−   gas jet effusing out of the pipe is also penetrated by the system of the oblique shock waves and is terminated by the mass flow cone and the dispersion zone.

**The sixth structure** ($\Delta p \cong 2.8 \times 10^4$ kgf/m$^2$):
−   inflow zone in the pipe looks like inflow zone of the first structure, but has bigger sizes and higher intensity of acceleration of gas to the inlet section of the pipe;
−   gas stream in the pipe has relative velocity M=1 in the inflow section; the normal shock wave appears in this section and the next motion of a subsonic gas stream in the pipe goes off with the frictional self-acceleration but remains subsonic;
−   gas jet effusing out of the pipe looks like the second structure.

**The seventh structure** ($\Delta p \cong 3.1 \times 10^4$ kgf/m$^2$):
−   inflow zone in the pipe maintains the spherical shape;
−   gas stream motion in the pipe is accompanied by its separation from the wall in front of the inlet into the pipe and in the entrance part of it forming entrance hydrodynamic vortical sublayer; the stream kernel accelerates to the supersonic velocity within limits of this sublayer and then the gas stream motion is penetrated by system of the oblique shock waves in the rest part of the pipe;
−   gas jet effusing out of the pipe is formed by the oblique shock waves and terminated by the mass flow cone and the dispersion zone of the flow rate.

**The eighth structure** ($\Delta p \cong 5.5 \times 10^4$ kgf/m$^2$):
this motion structure is the same as the previous but zone of the stream kernel separation from the inlet into the pipe gets maximum.

**The ninth structure** ($\Delta p \cong 9 \times 10^4$ kgf/m$^2$):
this structure of the motion looks like the seventh structure.

**The tenth structure** ($\Delta p \cong 10 \times 10^4$ kgf/m$^2$):
this structure of the motion looks like the sixth one.

**The eleventh structure** ($\Delta p \cong 12 \times 10^4$ kgf/m$^2$):
this structure of the motion looks like to the fifth one.

**The twelfth structure** ($\Delta p \cong (12 \ldots 38) \times 10^4$ kgf/m$^2$):
−   inflow zone in the pipe looks like inflow zone of the third structure;

– gas stream motion in the pipe is accompanied by the flow separation within limits of the pipe length forming the stream kernel and vortical hydrodynamic sublayer closed to the pipe wall; the nature of the gas flow in the kernel is rather close to isentropic; the stream is continuously accelerating to the supersonic velocity on all the pipe length; the M=1 section is shifting to the downstream as the pressure is increasing;

– gas jet effusing out of the pipe is formed by the oblique shock waves and is terminated by the mass flow cone and the dispersion zone of the flow rate.

**The thirteenth structure** ($\Delta p \cong (39...109) \times 10^4$ kgf/m$^2$):

– inflow zone in the pipe looks like inflow zone of the third structure;

– gas stream motion in the pipe is accompanied by the wall sublayer going out of the pipe outlet section limits; the M=1 section is slowly shifting to the pipe outlet section as the pressure is increasing;

– gas jet effusing out of the pipe is formed by the oblique shock waves and is surrounded by the transonic superlayer.

**The fourteenth structure** ($\Delta p \cong 111 \times 10^4$ kgf/m$^2$):

– inflow zone in the pipe maintains the half-spherical shape and has the least size comparing with previous structures;

– gas stream motion in the pipe is going on with frictional self-acceleration but remains appreciably subsonic within the pipe limits;

– according to the computation gas jet effusing out of the pipe contains the expanding part consisting of a transonic and supersonic zones and transonic superlayer enveloping the expanding part of the jet; this expanding part of the jet is terminated by the mass flow cone and the dispersion zone of the flow rate; interaction of the transonic superlayer with the expanding part of the jet results in overexpansion and shortening the latter; as a result, between the overexpanded part of the jet and its mass flow cone the pair of compression and expansion oblique shock waves appears. The expanding sonic and supersonic part of the jet ("j$_{sp}$" on fig. 1) is zone of the molecular self-acceleration of the gas jet in atmosphere.

**The fifteenth structure** ($\Delta p \cong 120 \times 10^4$ kgf/m$^2$):

The difference of this motion structure is that between the expanding part of the jet and its mass flow cone the system of several oblique shock waves can be placed.

Such fifteen structures described here are actual for the pipe with $L$=9$D$ and are possible for other types. The example of partial realization of possible motion structures is adduced in the third chapter of this work.

The experimental research of the structures of gas flow through the pipes and the convergent nozzles is connected with the certain technical difficulties, therefore only some examples of the visualizations of the flows and their diagrammatic representation can be found in special literature. For example, the results of rendition of the steam outflow through the convergent nozzle with the circular profile are submitted on fig. 11-15 in J.H.Keenan's book [10]. The mentioned images correspond to the fourteenth and fifteenth structures of the gas stream motion rotined on fig. 1 and circumscribed above in this paper.

The flow schemes of the gas stream in the outlet section of the pipe and the smoothly convergent nozzle built according to the results of an experimental research in Moscow Energy Research Institute and Central Boiler-Turbine Research Institute are presented on fig. 7-9 and 8-2 in E.M.Deich's book [11]. These schemes confirm a presence of the gas stream kernel and hydrodynamic sublayer both in the pipe and the smoothly convergent nozzle. Moreover, the scheme on fig. 7-9 shows that the critical section of stream (M=1) is inside of the pipe, not in its outlet section. The analogous feature is rotined on fig. 8-2.

VeriGas-program allows to determine structure of the gas stream motion in the pipes of any length. In particular, the pipe by relative length $L/D = 93$ is remarkable because of the fourteenth and fifteenth structures of the motion implemented just after the second structure's coming passing by ten intermediate structures from the third to the thirteenth one under monotonic increasing of pressure $p_0$ .

## 2. Flow characteristics of pipes and the circular profile and conical converged nozzles

The determination of the flow rate of the pipes and two types of the convergent nozzles as well is realized within the framework of the «General formulation» section. Thus the diameter of pipes is $D = 10$ mm $= 10^{-2}$ m. The computation of the metering characteristic of pipes is realized in a gamut of their relative length ($D/L$) from 0.6 up to 10,000. The results of determination of the mass flow rate of the pipes and the convergent nozzles are submitted on fig. 2–8. Relative length of the pipe is indicated on each diagram, for example, diagrams of the flow rate for the pipes with relative length 0.6; 1; 1.8; 2 are given on fig. 2.

The characteristic feature of the flow rate of pipes and the convergent nozzles is in a presence of the initial nonlinear segment and abrupt transition to the linear character. This transition corresponds to the beginning of the fourteenth structure of the gas flow (see fig. 1) through the flow element. The metering characteristic of the convergent nozzle with circular profile (see fig. 4) allows to explain an abnormal case, described in E.M.Tseyrov's work [9]. The author of this work has calculated the pressure change in vessel while its emptying through the convergent nozzle of circular profile and has taken the discharge coefficient $\mu = 1$. The calculation is realized by the author under condition of the isentropic outflow, which is disregarding of any losses. Then the author has conducted an experiment by emptying the tank under initial air pressure $16\times10^4$ kgf/m² into atmosphere and has done the calculation and experimental diagrams of pressure and time of emptying. The diagram indicates that the actual emptying of the tank in a gamut from initial pressure $16\times10^4$ up to ~ $8\times10^4$ kgf/m² is going quite faster, than calculated on an ideal adiabat. The curve CIRC0.6 on fig. 4 indicates that the mass flow rate corresponds to the fourteenth structure of the gas flow through the given flow element in a gamut of pressure from $16\times10^4$ up to $8\times10^4$ kgf/m² and essentially exceeds the isentropic mass flow rate. The flow rate drops almost halve under the pressure drop lower than $8\times10^4$ kgf/m² and then is non-linearly descending under further diminution of pressure drop. According to fig. 4, such character of change of the flow rate also allows to understand why the curve of the actual emptying on the Tseyrov's diagram under the tank pressure lower than $8\times10^4$ kgf/m² demonstrates essential diminution of the flow rate in comparison with isentropic calculation.

Some diagrams on fig. 6–8 have "downfalls". They are explained by flow transition from laminar to turbulent regime and back under monotonic increasing of pressure at inlet into the pipe.

The diagrams on fig. 7, 8 show that the flow rate curve becomes practically horizontal when the relative pipe length is $105D$ and more. It says that increase in the flow rate for long pipes including the gas pipelines can be got in general by increase in their diameters.

## 3. Solving of the inverse problem of gasdynamics

The inverse problems can also be solved with the help of VeriGas program based on the new qualitative approach and the evolutional mathematical model. For example, the results of photos of four structures of air jet motion out of the convergent axisymmetric nozzle under the supercritical pressure drops are shown on fig. 8-30 in the book mentioned above [11]. We put the task to know the geometrical parameters of the nozzle, to determine the state and motion parameters of the gas flow and the metering characteristic of this nozzle. Values of relative pressure of air while outflowing into atmosphere: $p_h/p_0 = 0.51$; 0.412; 0.267; 0.05 are given on fig. 8-30 [11].  Jacking $p_h = 10^4$ kgf/m², we gain $p_0$ for each of four offered cases: $2.029\times10^4$ , $2.43\times10^4$ , $3.74\times10^4$ and $20\times10^4$ kgf/m².

We find the kind of the flow element for convergent axisymmetric nozzle by cut-and-try method. First, let's take circular profile of the nozzle with diameter of the outlet section $D = 10$ mm and relative length $L/D = 0.6$. Such a nozzle has the least resistance. The result of determination of the metering characteristic of this nozzle is given on fig. 4. According to this diagram, the fourteenth outflow structure appears under $p_0 = 8 \times 10^4$ kgf/m$^2$, that is much earlier than in the problem conditions.

Now we take convergent nozzle of conical form and search for the nozzle profile angle to its axis by series of computations. As a result of computations, we have discovered that the fourteenth outflow structure appears under $p_0 = 20 \times 10^4$ kgf/m$^2$ for convergent conical nozzle with the profile angle 16.15 circular degrees and the $1.2D$ relative length.

We determine air state and motion parameters for the nozzle of found type under pressure drops mentioned above. The results of computation are given on fig. 9−13. Thus the $a$, $b$, $c$, $d$ points on the fig. 9 diagram match to pressure $p_0$. The graph of the sound velocity is indicated by a dashed line on the diagrams of the gas stream velocity on fig. 10−13. The comparison of the results of computation with offered results shows the complete quantitative and qualitative correspondence.

**Discussion of results**

The results of the direct and inverse computations of gas stream in such flow elements as pipe, nozzle evidence about regular qualitative physical conceptions, founded on the comprehensive approach to exposition the fluid motion within the United Physics framework. The results also confirm regularity of the mathematical model of the gas flow, which detects specific structures of the gas flow and reflects their evolutional nature even under one-dimensional formulation. This work represents physically adequate and mathematically precise solution of the fundamental problem of gasdynamics for the first time and opens a perspective to solution of the more complex and general problems about fluid motion as well. And, what is more, the authors tested successful efficiency of the problem solution about the gas stream motion in pipes and the complex flow systems, including the stationary and nonstationary heat exchange.

The results of solving the inverse problem presented on fig. 9−13 in the third chapter of this paper, are an example of presentation of the gasdynamics computation results of a flow element or system. Any incompleteness of results in comparison with this example will evidence about unsatisfactory solution.

**Final remarks**

This work is executed initiatively and independently by the scientists of Physical-Technical Group within the framework of development of subject «Fluid Motion Physics».
This work is made not disgracing of elderly and eminent contemporaries for, but increasing of the great precursors inheritance and teaching of new generation of creators for.

---

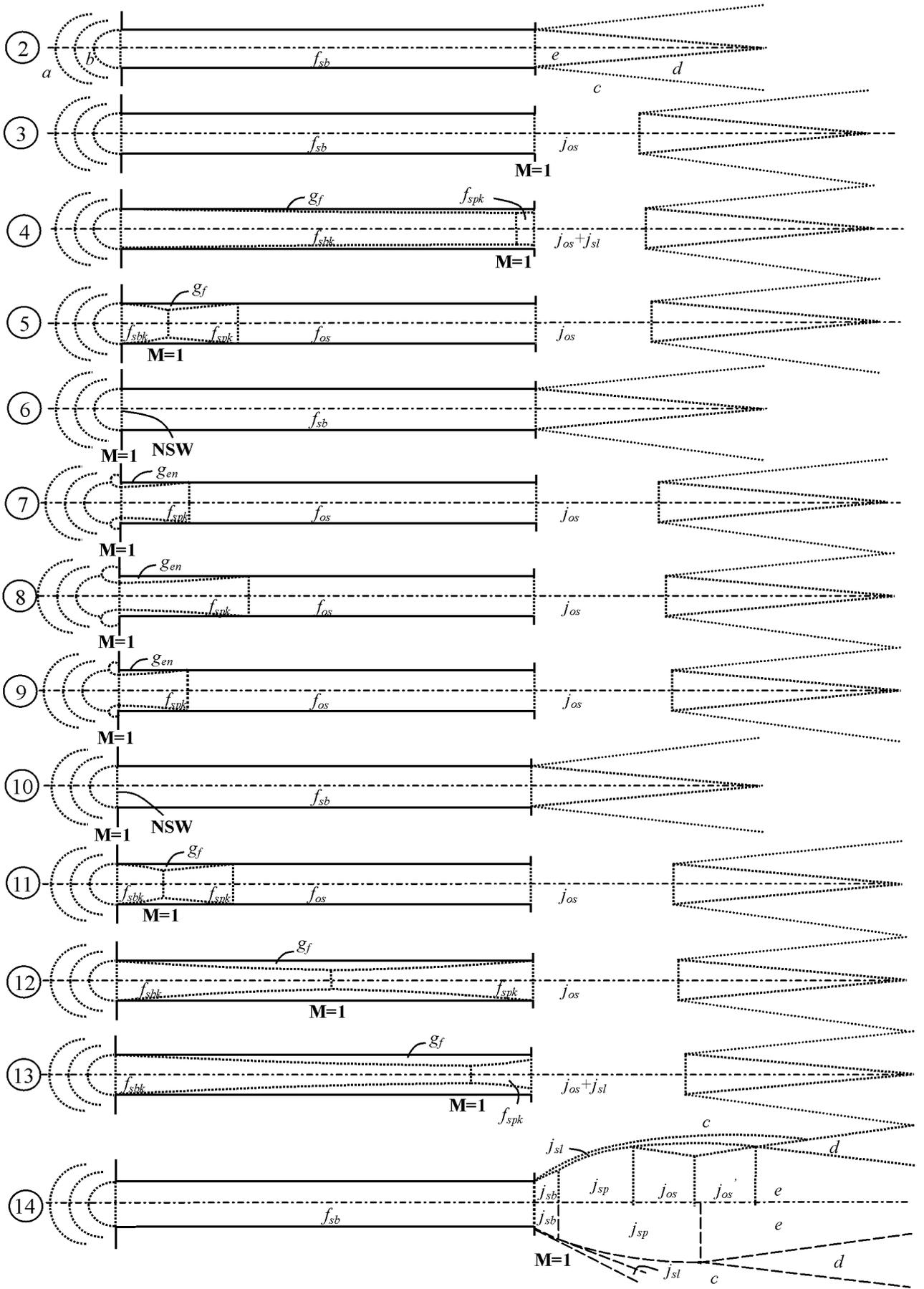

Fig. 1: Evolution of gas stream structures at flow through pipe
under action of the pressure drop increased from top to down

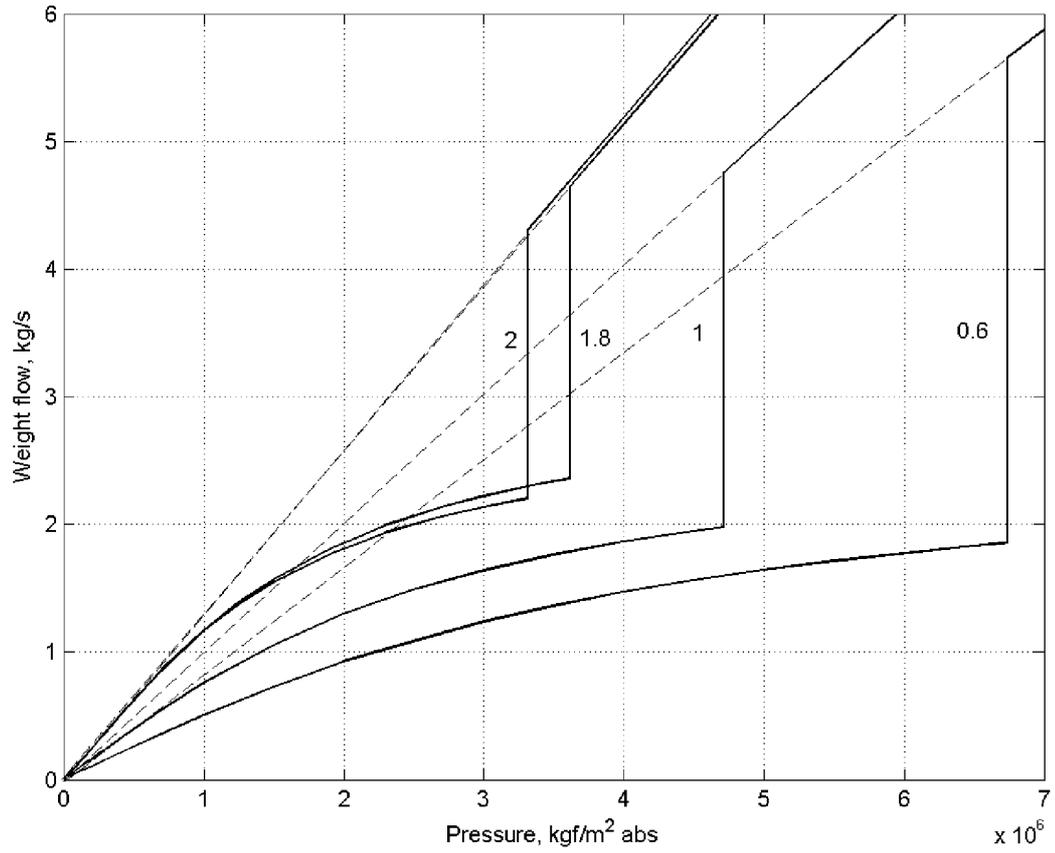

Fig. 2: Metering characteristics of very short pipes.

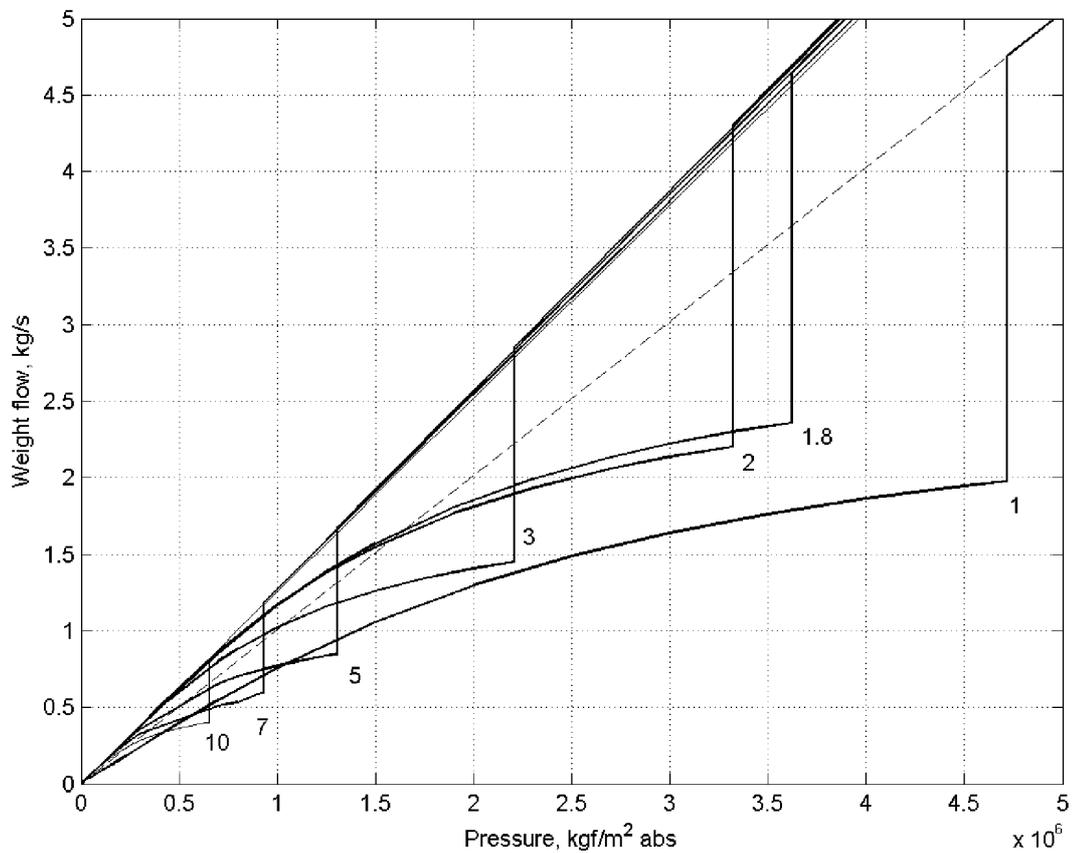

Fig. 3: Metering characteristics of short pipes.

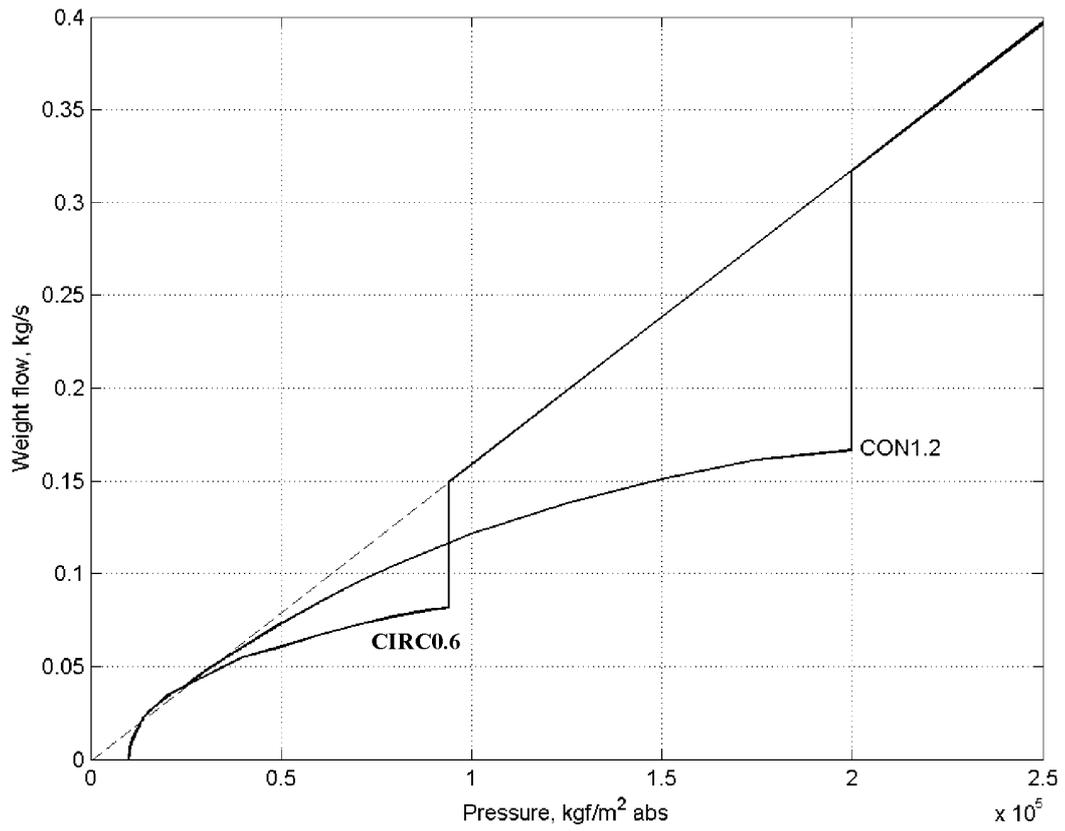

Fig. 4: Metering characteristics of circular profile and conical nozzles.

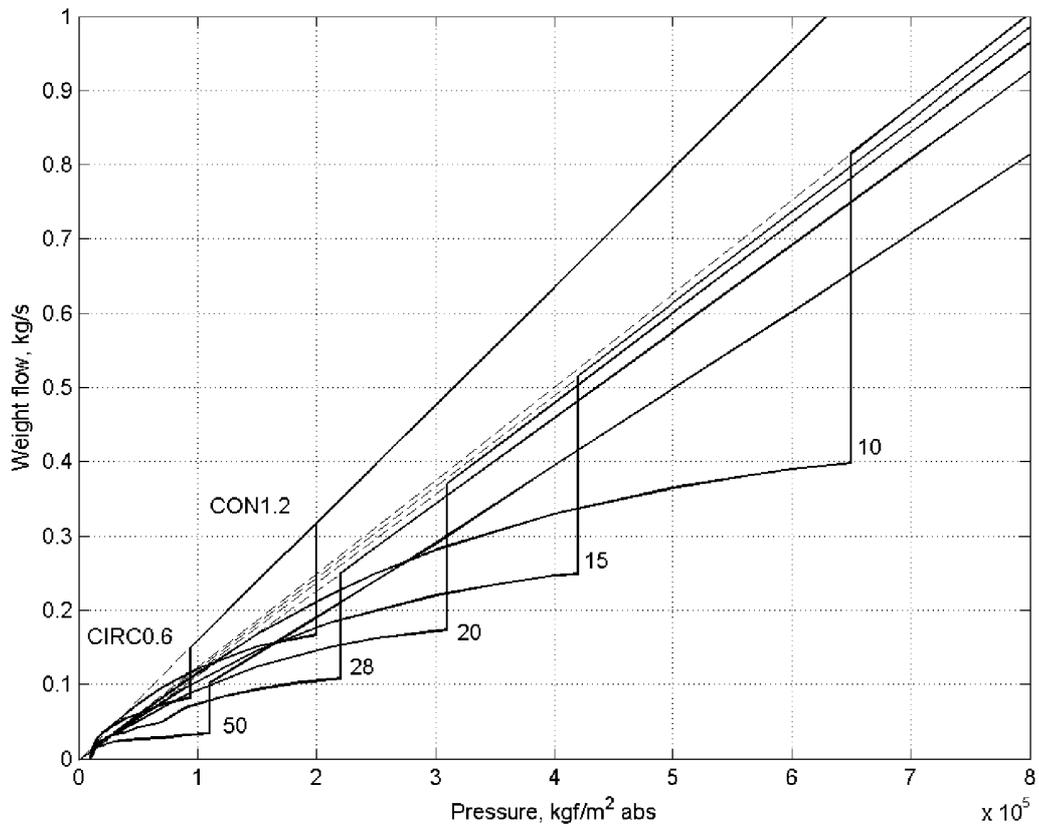

Fig. 5: Metering characteristics of pipes and the circular profile and conical nozzles.

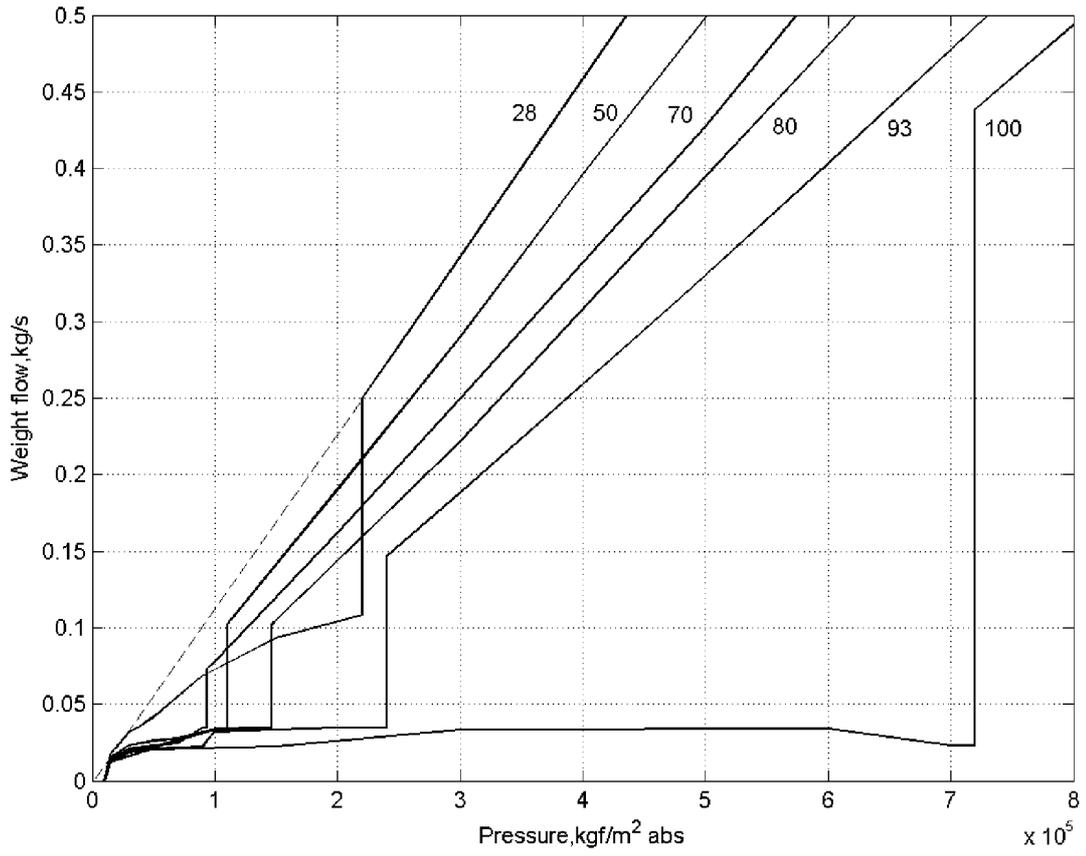

Fig. 6: Metering characteristics of pipes.

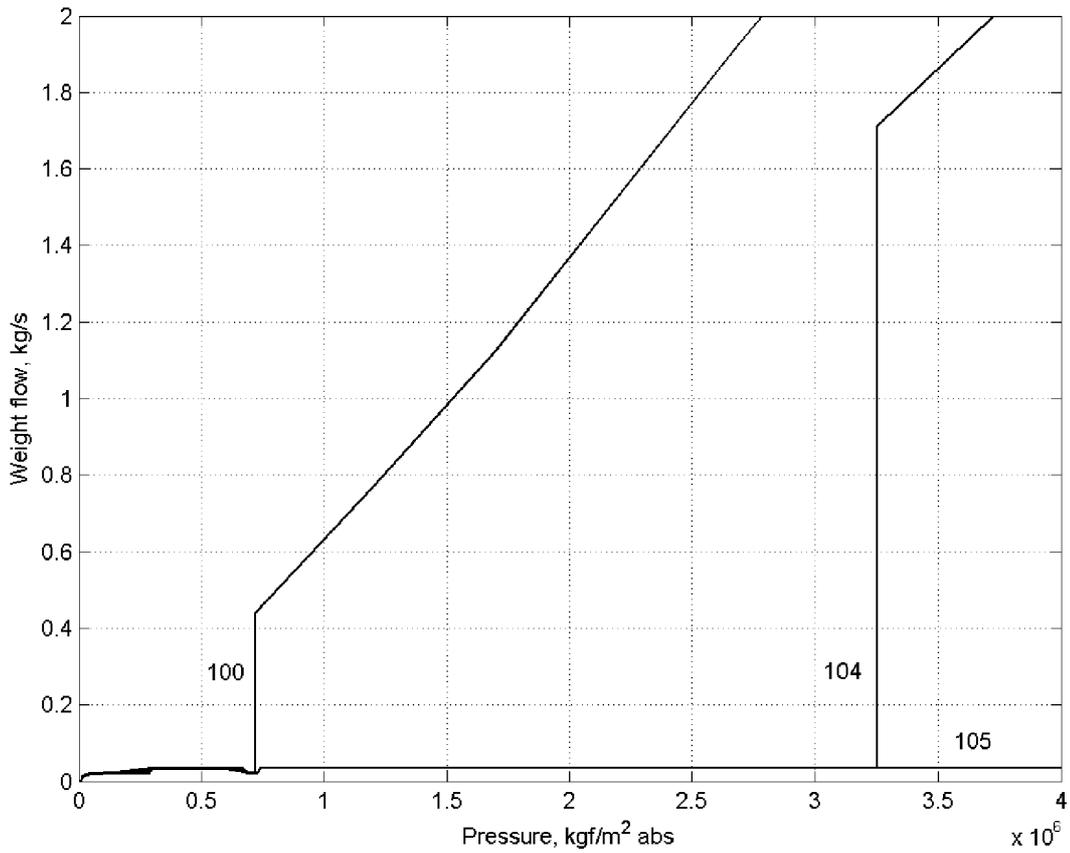

Fig. 7: Metering characteristics of pipes.

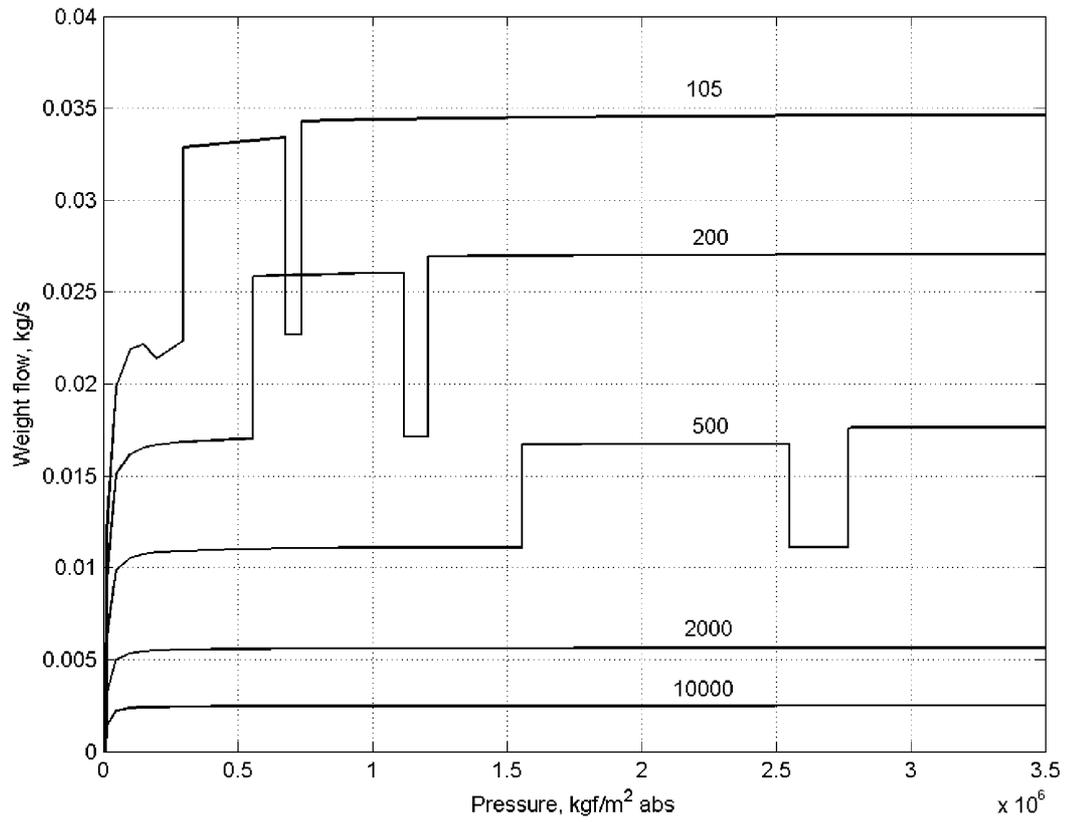

Fig. 8: Metering characteristics of long pipes.

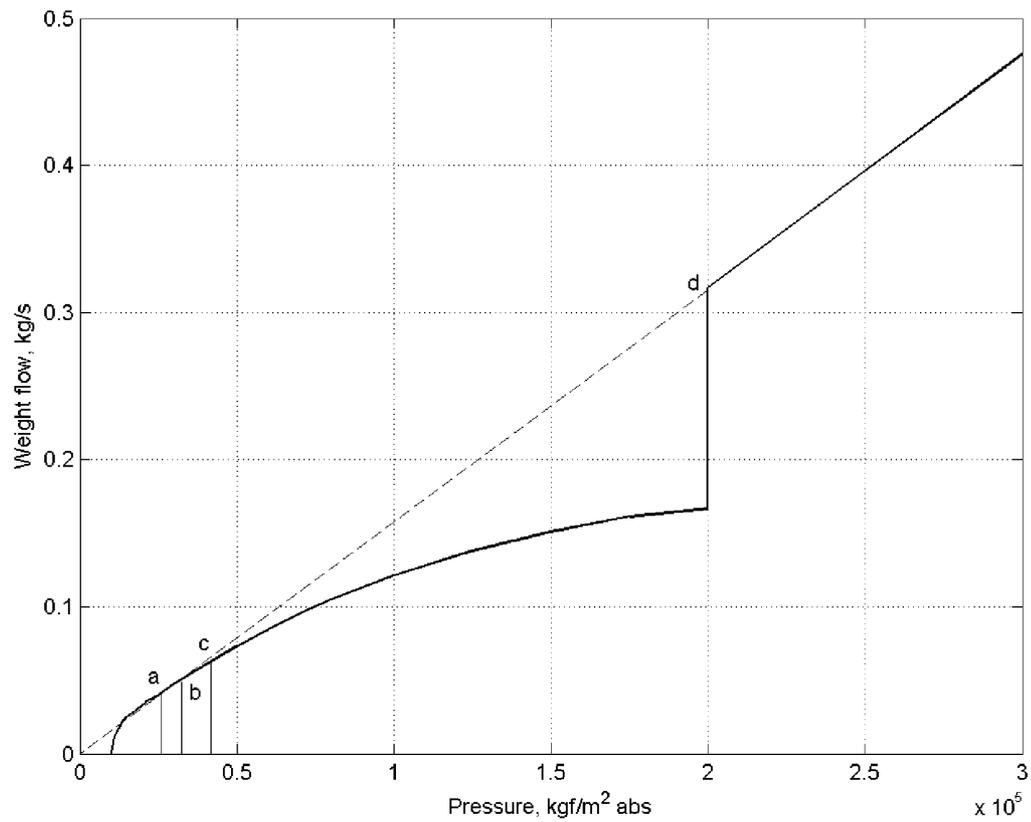

Fig. 9: Metering characteristic of conical nozzle.

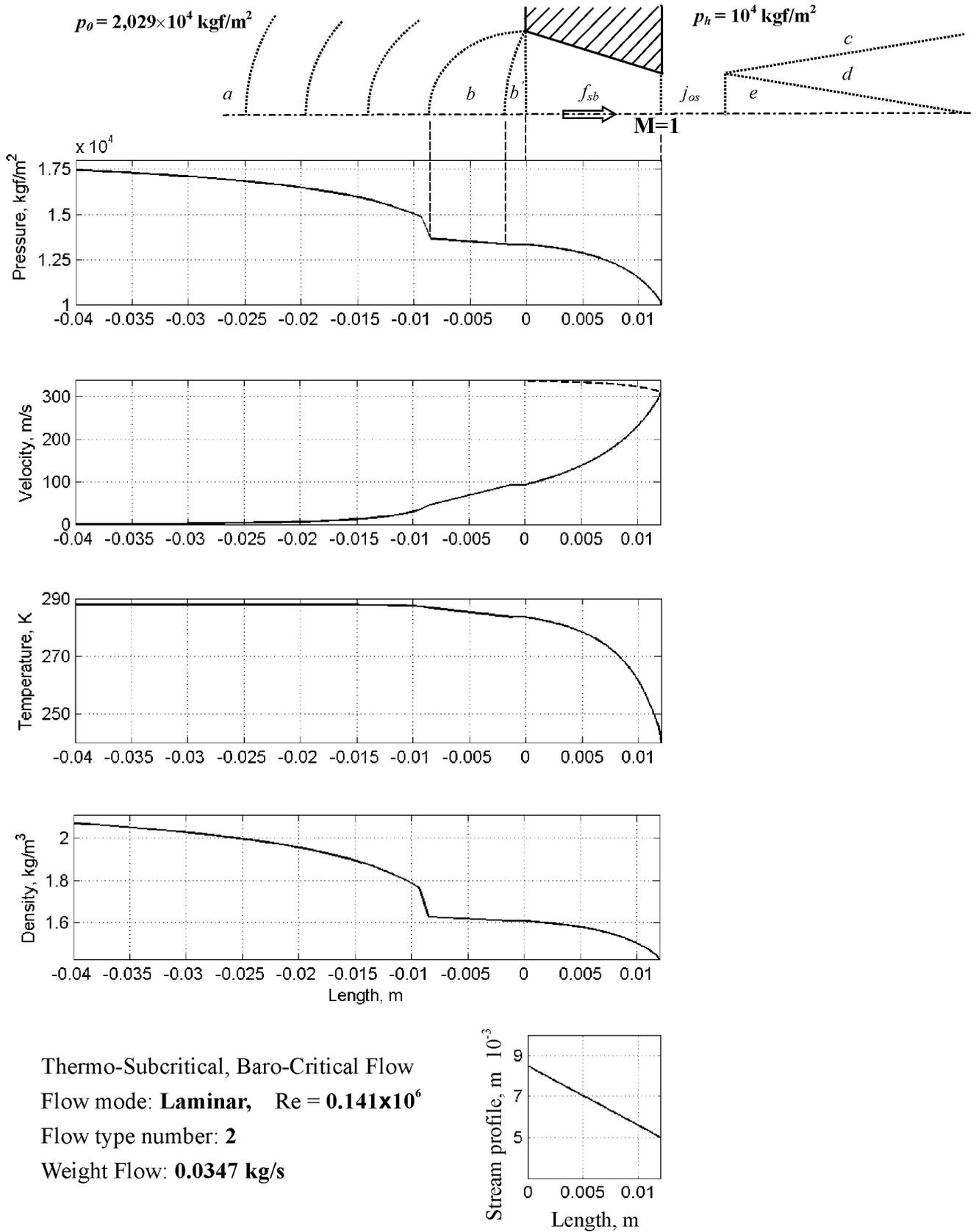

$p_0 = 2,029 \times 10^4 \text{ kgf/m}^2$

$p_h = 10^4 \text{ kgf/m}^2$

Thermo-Subcritical, Baro-Critical Flow

Flow mode: **Laminar,** Re = **0.141x10$^6$**

Flow type number: **2**

Weight Flow: **0.0347 kg/s**

Fig. 10: Stream parameters and structures at airflow through conical nozzle into atmosphere.

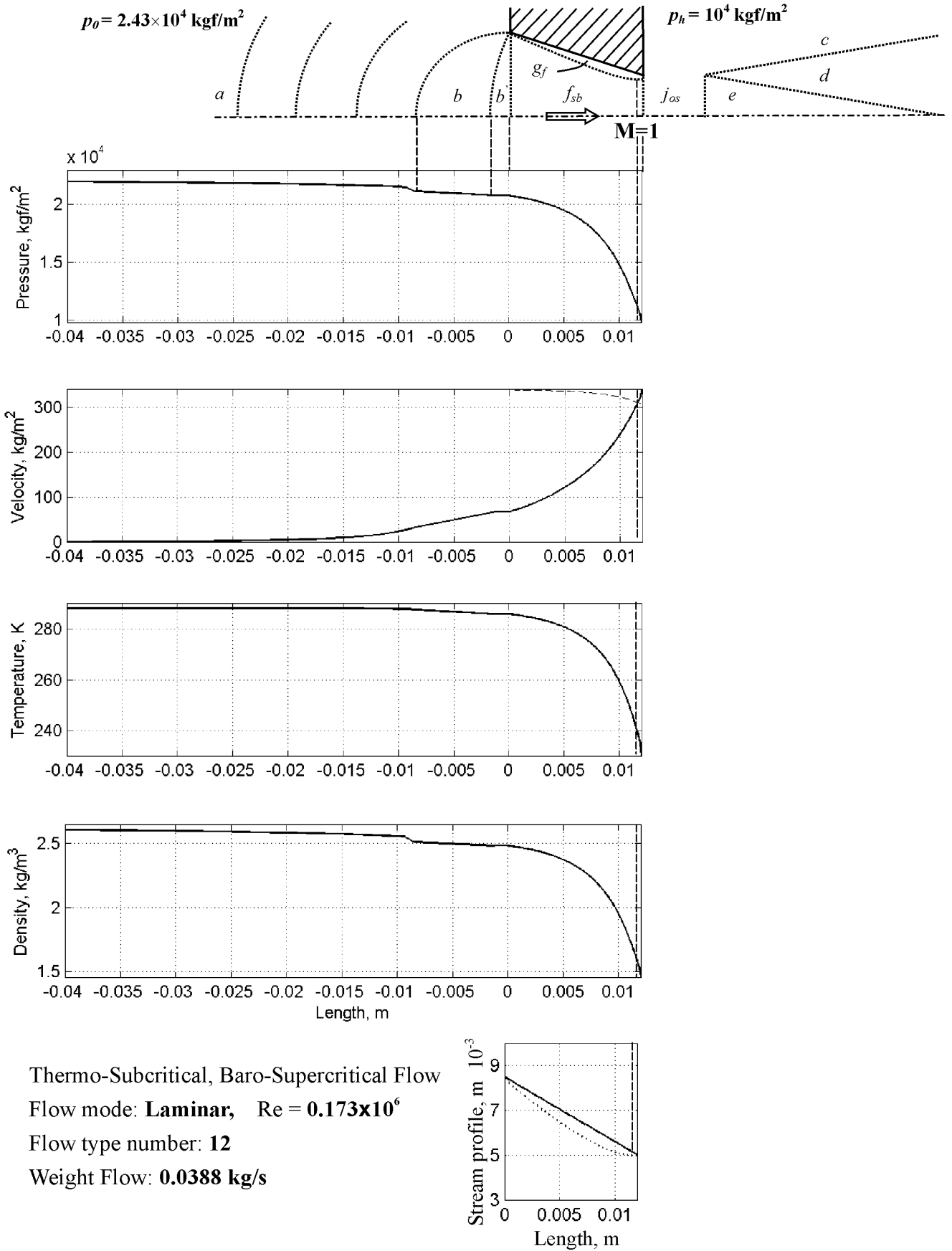

Thermo-Subcritical, Baro-Supercritical Flow

Flow mode: **Laminar,**   Re = **0.173x10⁶**

Flow type number: **12**

Weight Flow: **0.0388 kg/s**

Fig. 11: Stream parameters and structures at airflow through conical nozzle into atmosphere.

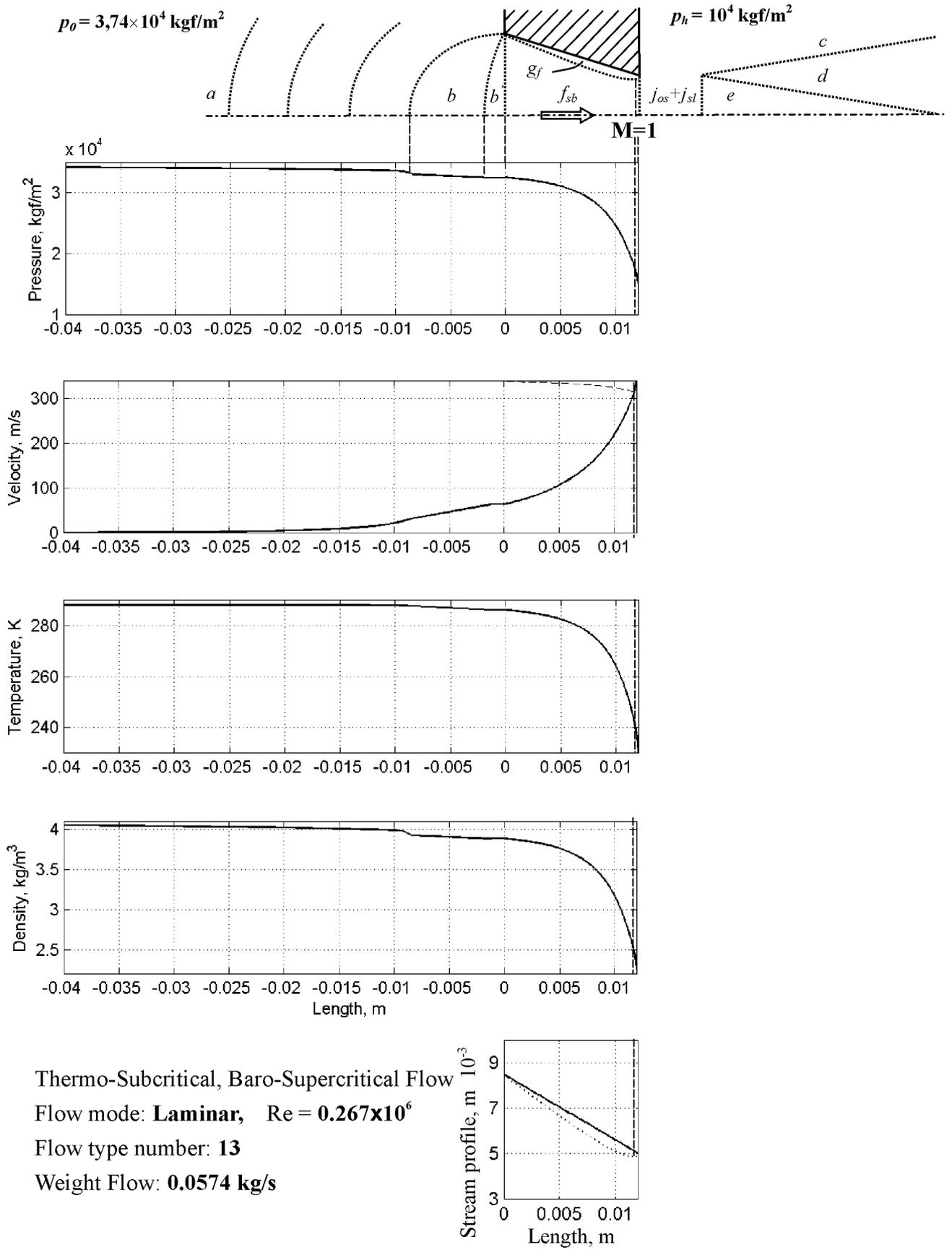

Fig. 12: Stream parameters and structures at airflow through conical nozzle into atmosphere.

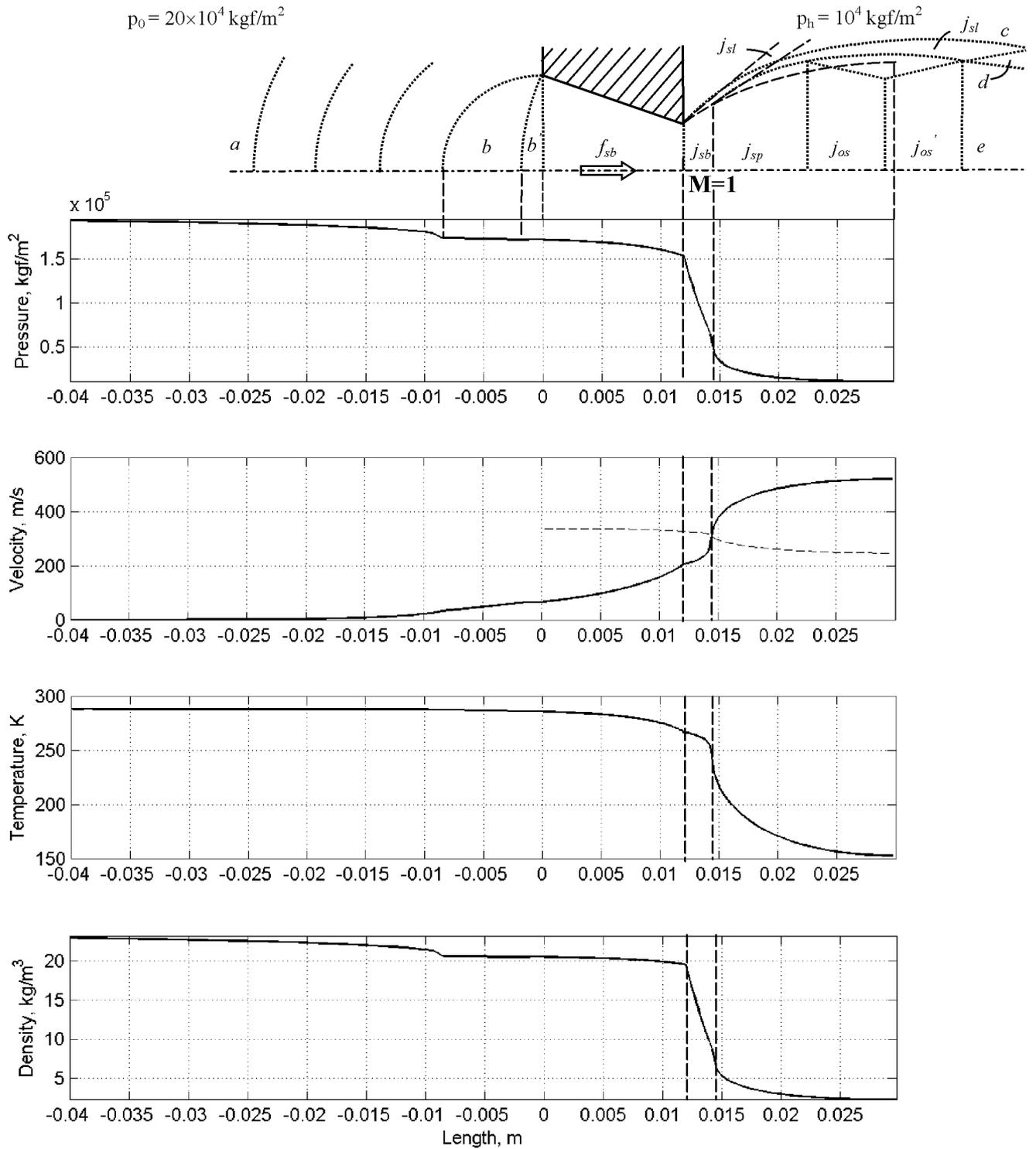

Thermo-Subcritical, Baro-Supercritical Flow

Flow mode: **Laminar,**   Re = **0.143x10^7**

Flow type number: **14**

Weight Flow: **0.267 + 0.050 = 0.317 kg/s**

   **(84.2 + 15.8   =   100%)**

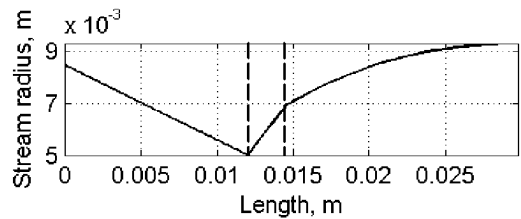

Fig. 13: Stream parameters and structures at airflow through conical nozzle into atmosphere.